# SEMIREGULAR VARIABLE STARS

## L. S. KUDASHKINA

The studies of semiregular variables of stars by different authors are considered, and the main theoretical and observational problems associated with these stars are reviewed. Their evolutionary status and possible connection with long-period variables such as Mira Ceti are discussed.

Individual objects belonging to different types of semiregular variables are described in detail.

After leaving the main sequence, the stars pass through the region of instability of Cepheids, turning into radially pulsating variables of type δ Cepheus. These stars can be associated with semi-regular variables giants and supergiants of spectral classes F – K, which are usually denoted by the symbol SRd. In the process of further evolution of the variables of high luminosity fall in the region of red supergiants, becoming the type variables SRc (or Lc), and the variables lower luminosity turn into a semiregular variables SRab (or wrong Lb) of late spectral classes.

Variables of the RV Tau type are a class of low–mass (with masses of the order of one solar) pulsating F – K − supergiants ($M_v = -3^m \div 5^m$), which may be at the short-term evolutionary stage of transition from the red giant to the protoplanetary nebula, which explains the small number of stars of this type of variability. Shklovsky (1956) was the first to point to stars of this type as the progenitors of planetary nebulae.

# ПОЛУПРАВИЛЬНЫЕ ПЕРЕМЕННЫЕ ЗВЕЗДЫ

## Л.С. КУДАШКИНА

Рассмотрены исследования полуправильных переменных звезд разными авторами, дан обзор основных теоретических и наблюдательных проблем, связанных с этими звездами. Обсуждается их эволюционный статус и возможная связь с долгопериодическими переменными типа Миры.

Подробно описаны отдельные объекты, принадлежащие разным типам полуправильных переменных.

Уйдя с главной последовательности, звезды проходят область нестабильности цефеид, превращаясь в радиально пульсирующие переменные типа δ Цефея. С этими звездами могут быть связаны полуправильные переменные гиганты и сверхгиганты спектральных классов F – K, которые принято обозначать символом SRd. В процессе дальнейшей эволюции переменные высокой светимости попадают в область красных сверхгигантов, превращаясь в переменные типа SRc (или Lc), а переменные меньшей светимости превращаются в полуправильные переменные SRab (или неправильные Lb) поздних спектральных классов.

Переменные типа RV Tau представляют собой класс маломассивных (с массами порядка одной солнечной) пульсирующих F–K – сверхгигантов ($M_v = -3^m \div 5^m$), находящихся, возможно, на кратковременной эволюционной стадии перехода от красного гиганта к протопланетарной туманности, что и объясняет немногочисленность звезд этого типа переменности. Шкловский (1956) первым указал на звезды этого типа, как прародителей планетарных туманностей.

1. *Введение.* Исторически к полуправильным переменным звездам относят пульсирующие переменные весьма различных типов. Например, желтые полуправильные звезды типа RV Тельца и красные полуправильные поздних типов, соседствующие на диаграмме Герцшпрунга – Рессела со звездами типа Миры Кита (рис. 1). При этом у первых кривая блеска достаточно регулярна, а у вторых поведение обладает такими неправильностями, что часто с трудом удается выделить какую-либо периодичность, чтобы классифицировать звезды. К тому же, красные полуправильные переменные от звезд типа Миры Кита часто отличает только меньшая амплитуда изменения блеска (менее 2.5 звездных величин).

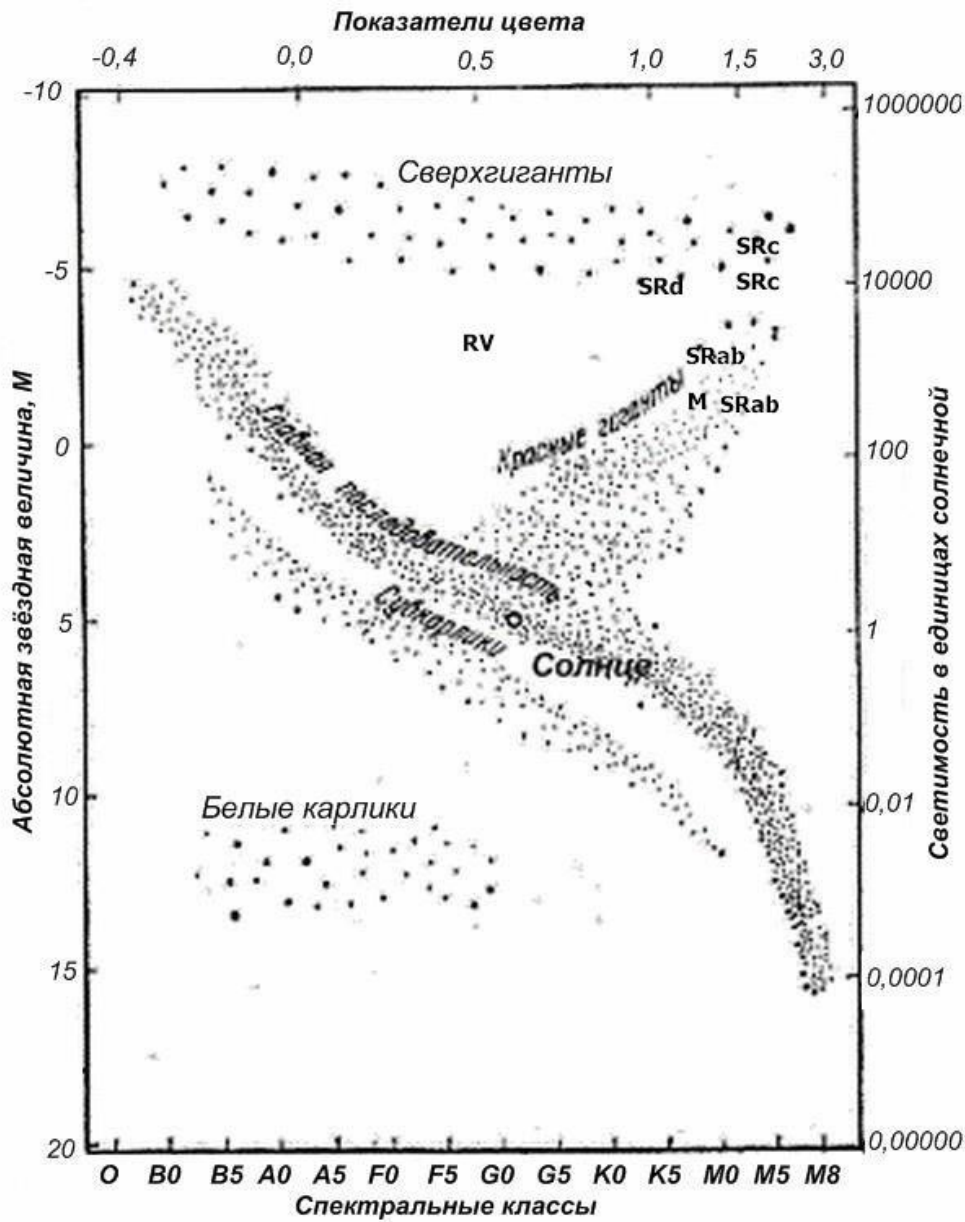

Рис. 1. Диаграмма Герцшпрунга – Рессела, на которой показаны приблизительные положения пульсирующих переменных: полуправильных сверхгигантов (SRc), полуправильных красных гигантов (SRab), долгопериодических типа Миры (M), полуправильных типа RV Тельца (RV) и полуправильных желтых гигантов и сверхгигантов (SRd). Одинаково обозначенные звезды, лежащие на диаграмме выше отличаются меньшим возрастом (см. табл. 4) [2, 3].

Все звезды асимптотической ветви являются пульсирующими.

Пульсационная неустойчивость возникает на определенных стадиях звездной эволюции, поэтому классификация пульсирующих переменных звезд по продолжительности периода, форме кривой блеска, виду спектра и другим наблюдательным признакам отражает их эволюционный статус, то есть принадлежность к группе звезд с определенными значениями массы, возраста и химического состава.

По сравнению со временем прохождения звездной стадии неустойчивости промежуток времени раскачки колебаний достаточно короток, поэтому практически все наблюдаемые нами пульсирующие звезды находятся на стадии автоколебаний.

В отличие от теории пульсаций классических пульсирующих звезд, которая была подробно разработана Жевакиным уже к середине XX века [4], детальной модели пульсаций долгопериодических переменных звезд (ДПП) пока не существует. Это связано с проблемами теории развития конвекции во времени, а также с определением мод пульсаций (что в свою очередь связано с неточностью определения радиусов, так как на поздней стадии эволюции звезд они имеют протяженную атмосферу, плавно переходящую в околозвездную оболочку). В результате многие наблюдательные проявления, такие, как скачкообразные изменения периодов, изменения амплитуды и формы кривой блеска не получили теоретического объяснения. Однако, некоторые наблюдательные эффекты, например, "прогрессивное" уменьшение периодов, получили приемлемое объяснение, или, по крайней мере, были высказаны гипотезы, хорошо согласующиеся с наблюдениями. Можно предположить, что развитие теории в существенной мере тормозится недостаточной классификацией наблюдаемых явлений. Например, мало изучено изменение кривых блеска от цикла к циклу, так как требует длинных рядов наблюдений и их тщательного анализа. Изучение закономерностей изменения кривых блеска от цикла к циклу, а также вековых изменений на протяжении длительных интервалов времени, а также их классификация могли бы стать наблюдательным фундаментом для развития теории пульсаций ДПП.

Основные решаемые в настоящее время проблемы, касающиеся звезд асимптотической ветви гигантов это – построение динамических моделей атмосфер, определение моды пульсации, изучение механизма потери массы и дальнейшая эволюция звезд.

Одним из главных вопросов теории звездных пульсаций является мода, в которой пульсируют звезды асимптотической ветви.

У каждой звезды существует свой набор периодов радиальных колебаний, который задается распределением вещества внутри звезды. Самый длинный из всех возможных периодов принадлежит колебаниям в фундаментальной моде. При пульсациях в первом обертоне внутри звезды имеется слой газа – узел обертона, который остается неподвижным на протяжении всего пульсационного цикла [5].

Для решения вопроса о типе звездной пульсации необходимо сначала классифицировать наблюдательные проявления звездной активности, то есть изменения блеска звезд.

Любые способы классификации звезд по типам переменности опираются на общий вид кривой блеска и спектральный класс. Однако, такой подход не всегда удачен, если речь идет о полуправильных переменных (SR). Для них часто нельзя рассматривать общую кривую блеска, так как она содержит участки, характерные для звезд различных типов. Происходит это, вероятно, потому что SR-звезды (от английского semiregular) в большинстве своем, во-первых, мультипериодичны и все компоненты этой мультипериодичности проявляют себя очень активно, то есть имеют сравнимую амплитуду с главным колебанием. А, во-вторых, период основного колебания также меняется [6].

2. *Данные.* Учитывая наблюдательные особенности полуправильных переменных (звезды яркие в визуальной области, спектры содержат молекулярные полосы, максимум энергии лежит в длинноволновой области, периоды несколько десятков или сотен дней), для фотометрических исследований удобно использовать наблюдательные базы ассоциаций любителей астрономии, такие как AFOEV (ftp://cdsarc.u-strasbg.fr/pub/afoev), VSOLJ (http://vsolj.cetus-net.org), AAVSO (http://aavso.org), ASAS (http://www.astrouw.edu.pl/asas). Данные наблюдений очень разнородны, поэтому обычно производится предварительная подготовка. Подробно о работе с базами данных наблюдений рассказывается в работе Андронова и Марсаковой [7] и Марсаковой и Андронова [8].

К указанным базам наблюдений можно добавить коллекции фотопластинок патрульных снимков неба, которые являются достаточно однородным наблюдательным материалом. Общее число накопленных фотопластинок составляет около 2.1 млн. Три обсерватории мира являются собственниками наиболее крупных фотографических коллекций (больше 100 тыс. астронегативов), прежде всего это Гарвардская обсерватория в США (500 тыс. негативов, с 1885 г.) и Зоннебергская обсерватория в Германии (около 300 тыс. негативов, с 1926 г.). В архиве Астрономической обсерватории Одесского национального университета им. И. И. Мечникова — третьей в мире обсерватории-собственника крупной фотографической коллекции — содержится около 104 тыс. пластинок, с 1909 г.: 20 тыс. старинных, включая коллекцию Симеизской обсерватории, и более 80 тыс., полученных в Одессе в с. Маяки, начиная с 1957 г. (большую часть этой коллекций составляют прямые фотографические снимки, выполненные по программам наблюдений переменных звезд) [9].

Исходя из теории пульсаций звезд, фундаментальной характеристикой, которую можно определить из наблюдений, является период изменения блеска. Для наиболее точного его определения необходимо иметь среднюю кривую блеска на большом интервале времени. Однако для полуправильных переменных это является весьма трудной задачей. Часто удобно использовать отдельные сезоны в некотором временном интервале, длительность которого заведомо превышает возможный период изменения блеска. Кроме того, можно исследовать период в разных диапазонах длин волн. Например, для поиска периода и построения средних кривых блеска полуправильной переменной $L_2$ Pup использовались наблюдения в ближних инфракрасных полосах H и K [10]. А для анализа средней кривой блеска слабой полуправильной переменной V411 Sct использовались данные [11].

3. *Методы.* Полуправильные переменные могут на отдельных интервалах показывать достаточно устойчивый период, который затем может смениться другим или плавно менять свое значение, а также колебания могут стать хаотическими. Периодограммный анализ позволяет определить значения циклов не только устойчивых колебаний, но и меняющихся со временем.

Для анализа используется метод наименьших квадратов, где сравнивается дисперсия отклонений от сглаживающей функции с дисперсией исходных наблюдений. В качестве тест-функции использована статистика

$$S(f) = \frac{\sigma_C^2}{\sigma_O^2} = 1 - \frac{\sigma_{O-C}^2}{\sigma_O^2}, \qquad (1)$$

где $\sigma_O$ – среднеквадратичное отклонение «наблюдений» $O$ от среднего. $C$ – соответствует «расчетным» значениям и $O$–$C$ отклонениям «наблюдаемых» значений от «расчетных». Основные уравнения, сглаживающие функции, определение моментов экстремумов методом дифференциальных поправок, а также одночастотное и многочастотное приближения гармоническими функциями подробно рассмотрены Андроновым [12] и в работе Андронова и Марсаковой [7].

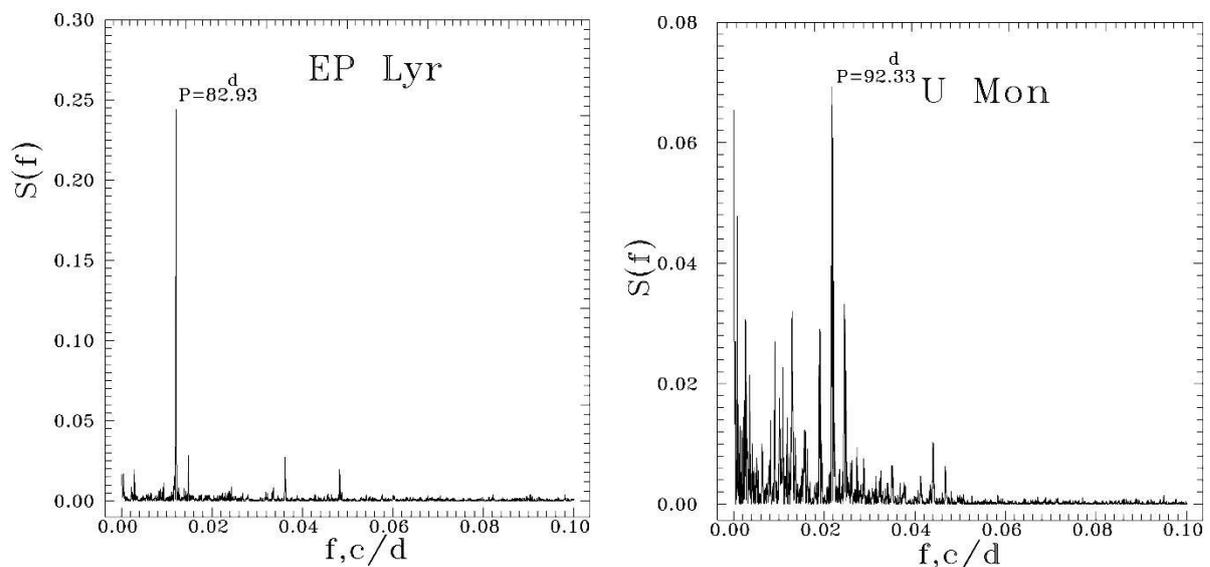

Рис. 2. Пример периодограмм для двух звезд типа RVb.

«Wavelet» или всплеск-анализ применяется для исследования процессов произвольной природы, в данном случае, для квазициклических, мультипериодических или хаотических колебаний блеска полуправильных переменных звезд. Как и в периодограммном анализе, алгоритм всплеск-анализа основан на методе наименьших квадратов с дополнительными весами. Здесь используется тест-функция, которая аппроксимируется синусоидой (для медленно меняющихся «периодов»), или скользящими синусами (если период постоянный). В отличие от периодограммного анализа, в данном методе определяется зависимость характеристик переменности от времени, например, периода и амплитуды.

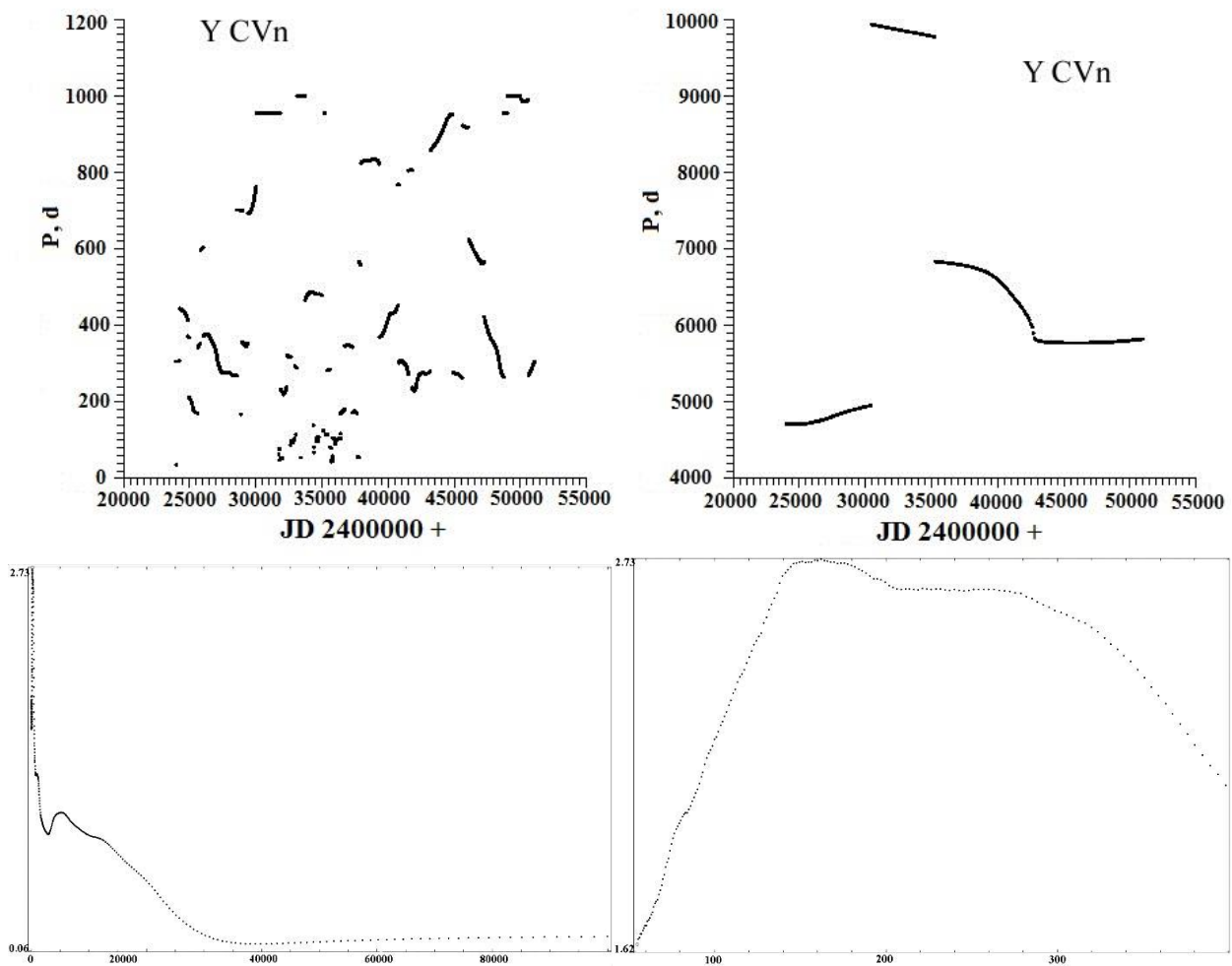

Рис. 3. Вплеск-анализ (вверху) и шкалограмма (зависимость тест-функции от пробного периода) для SRb – звезды Y CVn (внизу).

В случае если период меняется скачкообразно, либо колебания хаотичны, проводится шкалограммный анализ (зависимость тест-функции от шкалы – пробного периода), где тест-функция аппроксимируется параболой. Циклические или квазипериодические колебания приводят к появлению на шкалограмме отдельных пиков, хаотические колебания характеризуются отсутствием таких пиков и показывают только «непрерывный спектр». Основные соотношения, обсуждения разных модификаций метода всплеск-анализа в применении к переменности звезд разных типов, а также обсуждение оптимального выбора сетки пробных частот (периодов) можно найти в работе Андронова [13].

Еще один метод исследования, который можно использовать как дополнение к предыдущим, – анализ фазовых плоскостей (портретов). В

качестве координат фазовой плоскости $x$ и $\dot{x}$ используются $m$ – блеск звезды и его производная по фазе [14].

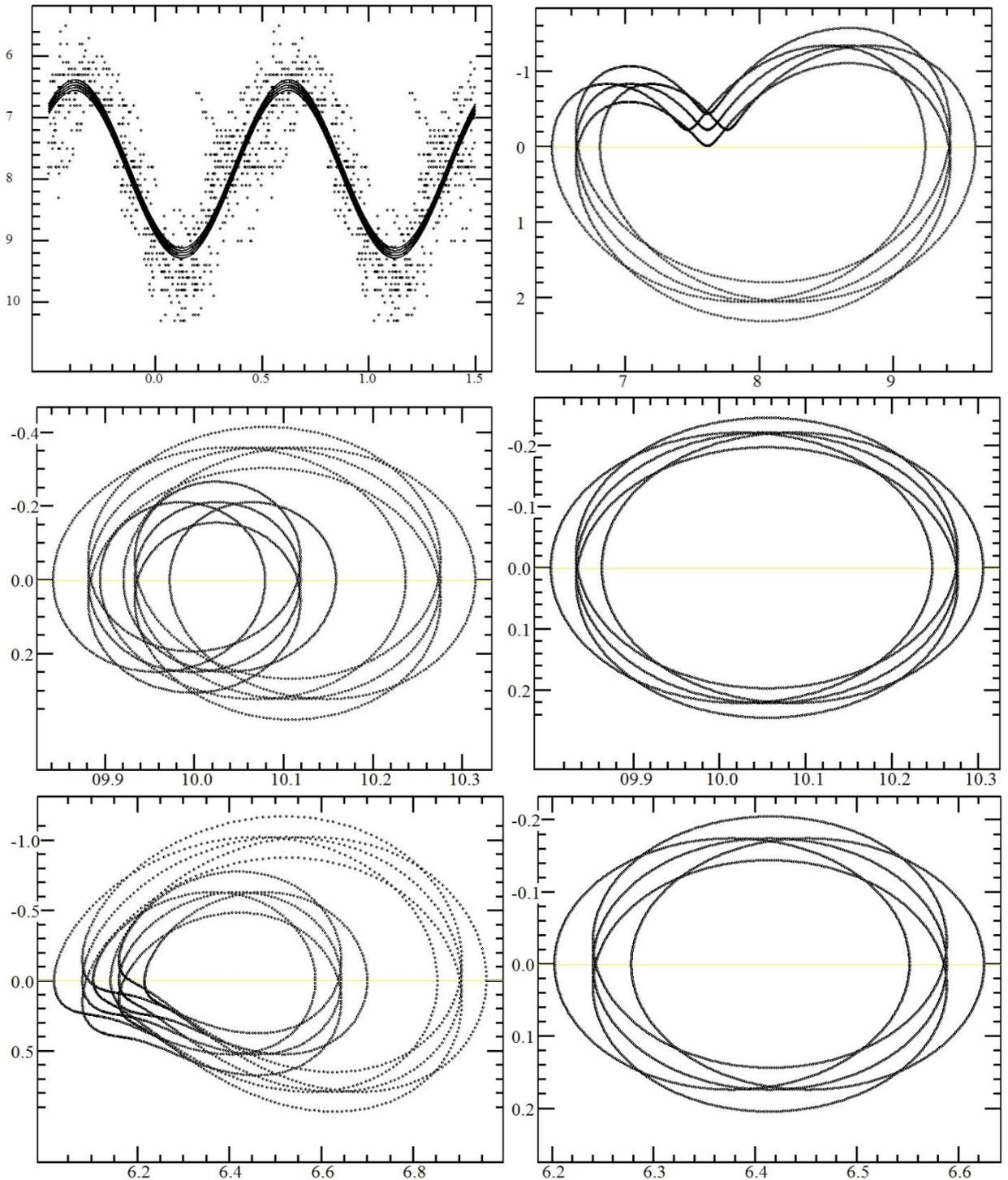

Рис. 4. Средняя кривая блеска [15] и фазовый портрет SRa – звезды W Hya [14] – вверху; фазовые портреты RV Tau с основным периодом (слева) и половинным (справа) – посередине и то же для U Mon – внизу [33].

Заметим, что кривые блеска звезд типа RV Tau, хотя и подвержены сильной изменчивости, но их фазовые портреты указывают на то, что колебания с половинным от основного (см. табл. 1) периодом лучше отражают предельный цикл автоколебательного пульсационного процесса.

4. *Объекты.* Характеристики некоторых полуправильных переменных звезд, которые будут рассмотрены, приведены в таблице 1 (данные взяты из ОКПЗ [37]).

*Таблица 1*

ХАРАКТЕРИСТИКИ НЕКОТОРЫХ ПОЛУПРАВИЛЬНЫХ ПЕРЕМЕННЫХ ЗВЕЗД (ОКПЗ)

| Звезда | Тип | Sp | P, дни | Δmag | Примечание |
|--------|-----|-----|--------|------|------------|
| W Hya | SRa | M7.5e – M9ep | 361 | $3.9^m$ | Δmag и форма кривой блеска сильно меняются |
| PZ Cas | SRc | M3Ia | 850 | 2.0 | Второй период $3195^d$ |
| S Per | SRc | M3Iae – M7 | 822 | 4.1 | Вторичные волны, мультипериодичность |
| Y CVn | SRb | C5,4J(N3) | 3000 | 2.46 | Мультипериодичность: $273^d$, $160^d$ |
| RX Boo | SRb | M6.5e – M8IIIe | 162.3 | 2.67 | $P_2=304.7^d$, $P_3=2692^d$ |
| RT Vir | SRb | M8III | 155 | 1.29 | |
| SV Peg | SRb | M7 | 144.6 | 1.8 | |
| TW Peg | SRb | M6 – M8 | 929.3 | 0.9 | Накладываются малые колебания с $P=94.13^d$ |
| BK Vir | SRb | M7III | 150 | 1.52 | |
| $L_2$ Pup | SRb | M5IIIe – M6IIIe | 140.6 | 3.6 | |
| AF Cyg | SRb | M5e – M7 | 92.5 | 2.0 | Накладывается более двух независимых колебаний $P_2=175.8$ и $P_3=942.2$ дней |
| U Mon | RVb | F8eVIb – K0pIb(M2) | 91.32 | 2.7 | Средняя величина меняется с периодом $P=2320^d$ |
| DF Cyg | RVb | G5 - K4I-II | 49.88 | 4.4 | Средний блеск меняется с периодом 780.2 дней |
| R Sct | RVa | G0Iae – K2p(M3)Ibe | 146.5 | 4.4 | [Fe/H]= -1. Период переменный |
| R Sge | RVb | G0Ib – G8Ib | 70.770 | 2.4 | 70.07<P<71.24, средний блеск меняется с периодом $1112^d$ |
| EP Lyr | RVb | A4Ib – G5p | 83.34 | 0.94 | Вторичные колебания $P_2=45.1$, $P_3=7067$ дней |
| RV Tau | RVb | G2eIa – M2Ia | 78.731 | 3.5 | Средний блеск меняется периодом $1224^d$ |

*W Гидры.* Поиск периода проводился программой FO [16], а затем периоды, соответствующие самым высоким пикам, уточнялись методом дифференциальных поправок программой Four-M [12].

Найдено три значения периодов $P_1=381.7^d\pm0.3^d$, $P_2=400.2^d\pm0.7^d$, $P_3=357.2^d\pm0.8^d$, соответствующие амплитуды $A_1=1.30^m\pm0.030^m$, $A_2=0.50^m\pm0.029^m$, $A_3=0.32^m\pm0.03^m$ и начальные эпохи для максимумов $T_{max1}=2446448.8\pm1.4$, $T_{max2}=2446506.9\pm3.7$, $T_{max3}=2446286.9\pm5.3$. Средняя кривая блеска с периодом $P_1=381.7^d$ показана на рис. 4.

*PZ Кассиопеи*. Эта SRa - звезда исследовалась нами на протяжении 14 лет [17]. Период этой звезды, видимо, имеет тенденцию к увеличению, так как первоначально определялся как 801 день, затем 830-842$^d$ и с момента JD 2448400 по 2449500 (1 полный цикл колебаний) определяется как 905 дней. Однако после этого вскоре звезда вступила в интервал постоянства блеска (с 1994 г.) и до начала 1997 г. показывала лишь незначительные хаотические поярчания. Наблюдения Брюханова [18] и Neumann [19] на участке JD 2448600 – 2449200 полностью совпадают (не считая небольшого систематического сдвига в звездных величинах).

У этой звезды возможно также наличие колебания с периодом около 300 дней, однако, это может также оказаться ошибкой селекции наблюдений.

*S Персея*. Эта звезда-сверхгигант SRc отнесена к звездам с гармонической переменностью. Период определен P=809.$^d$91 [20]. На самом деле на периодограмме присутствует не один пик а два. Первый соответствует периоду P=16173$^d$±158$^d$, с которым меняется средний блеск (заметим, что интервал наблюдений составляет более 20000 дней), а второй пик является сдвоенным $P_1$=809.6$^d$ ±0.22$^d$ и $P_2$=768.8$^d$ ±0.31$^d$.

Принимая, что звезда пульсирует в фундаментальной моде и, используя зависимость "период - абсолютная болометрическая величина", полученную Фистом [21], а также параметры, которые получил Абрамян [22] для сверхгигантов, был оценен радиус звезды R = 1.0·10$^{14}$ см или примерно 1400$R_☉$. При этом пульсационная константа будет равна Q≈0.077, что, вообще говоря, согласуется с теоретическими значениями (Q от 0.06 до 0.08) для полуправильных переменных [23].

*Y Гончих Псов*. С 1981 по 2010 годы мы находим упоминания у разных авторов необычного поведения блеска этой звезды.

Эта яркая углеродная холодная звезда редкого J-типа имеет обособленную асимметричную оболочку. Звезда расположена не точно в центре, яркость оболочки в западной части ниже. Толщина оболочки (2-5)·10$^{17}$ см при внутреннем радиусе оболочки 7·10$^{17}$см. Расстояние до звезды 250 пк. Темп потери массы уменьшился за последние 14000 лет на два порядка. Похожие вариации встречаются у звезд U Hya, U Ant. Но Y CVn не показывает абсорбционную линию технеция. Утверждается, что у нее не идет s-процесс. Из-за этого Y CVn можно поместить не на AGB, а на RGB или на стадию стационарного горения гелия в ядре после гелиевой вспышки [24]. Звезда была исследована с помощью периодограммного и всплеск-анализа (рис. 3). В литературе встречаются весьма разнообразные значения периодов, которые сведены в таблицу 2.

*Таблица 2*

ЗНАЧЕНИЯ ПЕРИОДОВ ДЛЯ ЗВЕЗДЫ Y CVN ПО ИССЛЕДОВАНИЯМ РАЗНЫХ АВТОРОВ

| Автор, год | Значение периода | Работа |
|---|---|---|
| Krisciunas, 1981 | 187$^d$ с амплитудой 0.5$^m$ и 98$^d$ с амплитудой 0.15$^m$ | [26] |

| Vetešnik, 1983 | Min=2436097.5+251.8E | [27] |
|---|---|---|
| Barnbaum, 1992 | 157$^d$ | [25] |
| Kučinskas, 1992 | 4000$^d$ | [28] |

По нашим исследованиям наиболее характерными значениями периодов для этой звезды являются значения от 247$^d$ на интервале JD 24 45322-46499 до 343$^d$ на интервале JD 24 24362-27129 [29].

*RX Волопаса.* Исследование этой звезды типа SRb, проведенное Шаповаловой [20] по всему массиву данных AFOEV и VSOLJ, подтвердило наше значение периода около года (P=369$^d$). Однако, характерная мультипериодичность с периодами 162$^d$ и 179$^d$ [30] не была выявлена, хотя на периодограммах присутствуют пики в этой области (≈164 и 183 дня), но их отношение сигнал/шум низко.

*RT Девы.* Значение периода из ОКПЗ (табл. 1) противоречит некоторым наблюдательным данным. Например, Венцель [31] получил значение периода около 200 дней из кривой блеска. Также обнаружены признаки систематической переменности в радиодиапазоне с периодом также близким к 200 дням. Исследования периода в работах Андронова и др. [32, 33], показали, что период этой звезды меняется со временем и последнее значение найдено равным примерно 136 дней. Однако на периодограммах присутствует сразу несколько пиков, удовлетворительно описывающих среднюю кривую блеска. Учитывая неоднородность наблюдений, периодограммный анализ был проведен с использованием методики «сглаживания сглаживающих сплайнов» [34] с числом базисных функций $NF = 5$. В качестве тест-функции использовался коэффициент корреляции $r$ между вычисленными (с использованием «наилучшей» в смысле метода наименьших квадратов функции) и наблюдаемыми значениями для каждого пробного периода. В качестве критерия значимости пиков на периодограмме использовалось значение $\rho = r/\sigma_r$, где

$$\sigma_r = \sqrt{(1-r^2)/(N-NF)}, \qquad (2)$$

где $N$ – число наблюдений. По правило «трех сигма», значимыми пиками можно считать с достоверностью 99% те, для которых $\rho>3$.

В таблице 2 приведены значения периодов, соответствующих локальным максимумам на периодограмме, и оценки погрешностей их определения $\sigma_P$, вычисленные по формуле

$$\sigma_P = \frac{2\sigma_{O-C}^2(P_0)}{(N-NF)\left(\dfrac{d^2\sigma_{O-C}(P)}{dP^2}\right)_{P=P_0}}, \qquad (3)$$

где

$$\sigma_{O-C}^2(P) = \sigma_0^2\left(1 - r^2(P)\right), \qquad (4)$$

где $\sigma_0$ – среднеквадратичное отклонение наблюдений от среднего, $\sigma_{O-C}$ – от сглаживающей кривой, $P_0$ – «оптимальное» значение периода, соответствующее локальному пику.

*Таблица 3*

ЗНАЧЕНИЯ ПЕРИОДОВ В РАЗЛИЧНЫЕ ИНТЕРВАЛЫ ВРЕМЕНИ ДЛЯ ЗВЕЗДЫ RT VIR ПО НАБЛЮДЕНИЯМ РАЗНЫХ АВТОРОВ [33]

| P, d | 1955-1969 | $\sigma_P$ | 1957-1980 | $\sigma_P$ | 1970-1986 | $\sigma_P$ | 1979-1987 | $\sigma_P$ |
|---|---|---|---|---|---|---|---|---|
| 100-150 | | | 103.54±0.51 | 4.03 | 101.13±0.56 | 3.50 | | |
|  | 111.63±0.37 | 6.79 | 113.34±0.52 | 4.14 | 115.43±1.00 | 3.43 | 115.48±0.76 | 4.75 |
|  | 118.62±0.63 | 4.68 | 118.60±0.41 | 6.05 | | | | |
|  | | | 140.24±0.57 | 5.63 | 140.19±1.08 | 3.08 | 136.04±0.45 | 5.87 |
|  | | | 144.51±0.78 | 5.72 | | | 136.10±0.83 | 5.87 |
| 150-250 | 159.22±0.82 | 8.12 | | | | | 163.53±2.50 | 5.10 |
|  | | | | | 167.14±1.24 | 5.59 | 169.27±2.29 | 5.31 |
|  | | | 177.53±0.85 | 7.29 | | | | |
|  | 187.57±1.27 | 6.48 | | | | | | |
|  | | | | | | | 197.91±1.17 | 6.08 |
|  | | | 227.47±1.22 | 6.44 | | | 224.72±2.55 | 5.82 |
|  | | | 237.36±0.86 | 7.05 | | | | |

*SV Пегаса.* Тип SRb. Поиск периода по всему массиву наблюдений (примерно 100 лет) приводит к значениям $359.69^d \pm 2.46^d$ (отношение сигнал/шум ≈13); 764.84±17.35 (≈5.8); 12.82±0.01 (≈4.9). Более ранние исследования в двух цветах (где использовались более однородные наблюдения) показали наличие периодов $316^d$ и $170^d$ [35].

*TW Пегаса.* Тип SRb. Для исследования периодичности этой звезды использовались наблюдения разных авторов, в том числе и членов AFOEV. Данные были разделены в однородные группы, по которым и проводился поиск периода. Значение из ОКПЗ [42], равное 929.3 дней не подтверждено ни на одном из исследуемых интервалов. Более подробно об исследованиях этой звезды можно найти в работе [35].

*BK Девы.* Исследования периода позволило выявить два значения характерного времени переменности звезды: 147 дней, что близко к значению периода, указанного в ОКПЗ, но более уверенным нам кажется значение 264.30±0.47 дней. За время исследования блеск переменной менялся с амплитудой $0.75^m$ в фотовизуальных лучах (pv) [35].

*L$_2$ Кормы.* Исследование периода проводилось в полосах H и K по данным работы [10]. Определено значение периода близкое к указанному в ОКПЗ: H-полоса – 137.14±0.19 дней и K-полоса – 137.19±0.18 дней. Для анализа использовались программы Андронова [12, 16]. Период в 137 дней хорошо описывает среднюю кривую. Возможно также, что у звезды присутствуют колебания с циклом около 5511 дней [35].

*AF Лебедя.* Звезда AF Cyg вполне может быть прототипом отдельного класса звезд (как, например, RV Tau). Этот объект показывает последовательное "переключение" колебаний, значения периодов которых никак не зависят друг от друга [37]. Подобные ей звезды – RX UMa, W Cyg, RU Cyg [20].

*U Единорога.* Для изучения возможной мультипериодичности медленного колебания, предварительные пары частот были взяты из одночастотного приближения, и затем уточнены методом дифференциальных поправок. В результате обнаружено, что медленное колебание является бимодальным с периодами $P_1 = 2006 \pm 38^d$, $P_2 = 1382 \pm 70^d$ [35]. В то же время по нашим исследованиям у звезды наблюдается мультипериодичность с характерными для объекта типа RV Tau периодами $P=46^d.183 \pm 0.001$, $P=91^d.32 \pm 0.07$. На рис. 4 показаны фазовые портреты звезды с этими периодами.

На рис. 5 изображено изменение среднего блеска звезды U Mon, которое показывает квазипериодические колебания с амплитудой около $2^m$.

*DF Лебедя.* Для этой звезды было проведено исследование фазового портрета. В отличии от других звезд типа RV Tau, для DF Cyg более регулярный квазиэллипс получается для периода 776.4 дня, то есть с периодом, близким к периоду изменения среднего блеска, а не половинного от основного [38].

Для R Sct, R Sge, EP Lyr и RV Tau также изучались изменения среднего блеска, амплитуды и фазы максимума блеска со временем. В большинстве случаев не удалось выделить какой-либо период, так что эти колебания имеют скорее хаотический характер [39].

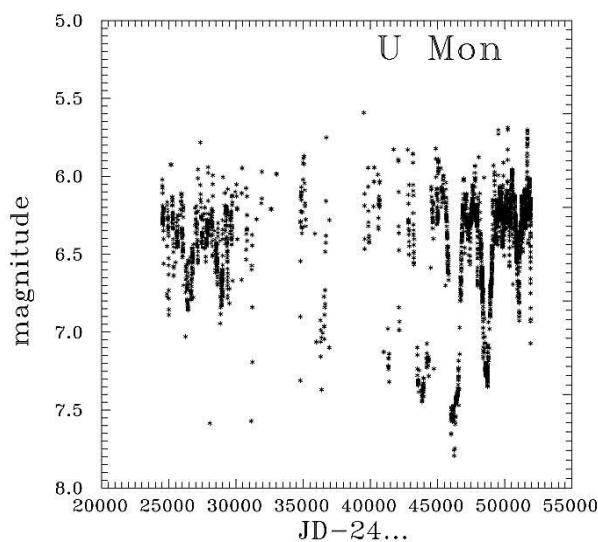 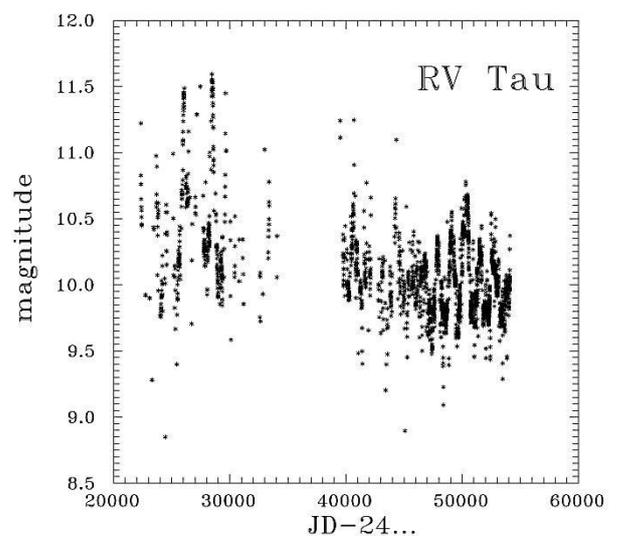

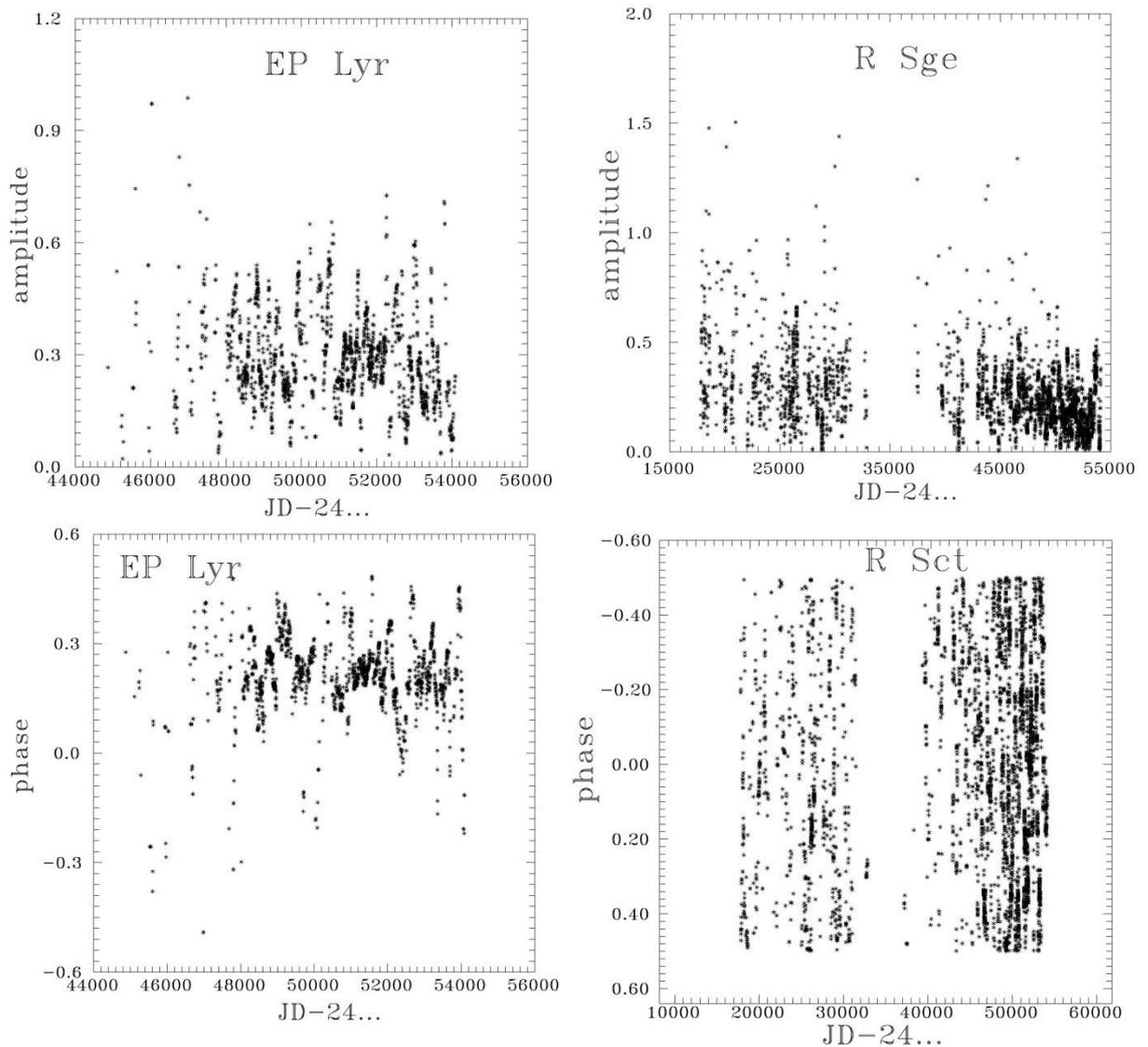

Рис. 5. Изменение среднего блеска – вверху; амплитуды – посередине; фазы максимума блеска – внизу – со временем для звезд типа RV Tau.

5. *Классификация.*

На сегодняшний день имеется очень грубая классификация SR-звезд, которую можно обобщить, используя три фундаментальные работы "Пульсирующие звезды" [40], "Общий каталог переменных звезд" [42], "Переменные звезды" [41], следующим образом:

*Таблица 4*

КЛАССИФИКАЦИЯ ПОЛУПРАВИЛЬНЫХ ПЕРЕМЕННЫХ ЗВЕЗД

| SRc | $0<t<10^7$ лет | поздние спектры (M) сверхгиганты амплитуды порядка $1^m$ периоды от 30 до нескольких тысяч дней | μ Cep |
|---|---|---|---|

| SRc | $10^7 < t < 10^9$ лет | M, C, S спектры<br>сверхгиганты | RS Cnc 1700$^d$ |
| --- | --- | --- | --- |
| SRb | $10^7 < t < 10^9$ лет | M, C, S спектры<br>гиганты<br>средний цикл от 20 до 2300 дней, возможно три вида поведения: квазипериодическое, постоянное, хаотическое | RR CrB<br>AF Cyg |
| SRb | $t > 10^9$ лет | M, C, S спектры<br>Маломассивные гиганты (меньше 1.3 массы Солнца)<br>Интенсивно теряют массу<br>Колебания блеска часто хаотичны | RT Vir<br>RX Boo |
| SRa | $10^7 < t < 10^9$ лет | M, C, S спектры<br>гиганты<br>амплитуды меньше $2.5^m$<br>периоды в пределах от 35 до 1200 дней<br>форма кривой блеска сильно меняется<br>имеют эмиссионные линии | очень похожи на звезды типа Миры Кита<br>Z Aqr 136.$^d$9<br>M1e-M3e |
| SRa | $t > 10^9$ лет | M, C, S спектры<br>Маломассивные гиганты (меньше 1.3 массы Солнца)<br>Интенсивно теряют массу | W Hya<br>VX Sgr |
| SRd | $10^7 < t < 10^9$ лет | F, G, K спектры<br>гиганты и сверхгиганты<br>Амплитуды от $0.01^m$ до $4^m$<br>Периоды от 30 до 1400 дней<br>отличаются от остальных отсутствием или очень слабыми полосами окиси титана, большими скоростями и светимостями | SX Her<br>SV Uma<br>UU Her<br>AG Aur |

Следует заметить, что SRc-класс фактически отмечает только звезды-сверхгиганты чаще всего с переменностью типа SRa.

Таким образом, обширный и неоднородный класс полуправильных переменных требует внимательного подхода и ревизии, которую вполне успешно можно провести, используя современные математические методы и уже имеющийся наблюдательный материал.

Прежде всего, полуправильные переменные звезды принадлежат AGB или post-AGB, то есть это в основном – красные гиганты и сверхгиганты.

Однако, встречаются объекты, которые, возможно, все еще находятся на стадии RGB (например, Y CVn).

Гоффмейстер [41] делит полуправильные на 4 группы – SRa, b, c, d, - а также, звезды типа RV Тельца. Такой же классификации придерживается и Общий каталог переменных звезд [42]. Однако имеются интересные работы, предлагающие дополнительное разделение полуправильных переменных. Например, согласно работам Кершбаума и Хрона [43, 44], в которой были изучены основные свойства SRa и SRb переменных, а именно – болометрические величины, периоды, амплитуды в визуальной и ИК областях, химические свойства, пульсационные свойства (диаграммы период – цвет, амплитуда – цвет, цвет – цвет) и пространнственно-кинематическое распределение, следует, что SRa звезды являются промежуточными объектами между миридами и SRb во всех аспектах исследования. SRb могут быть разделены на две группы соответственно присутствию или отсутствию эмиссии от околозвездной пыли. Это разделение также соответствует разделению по периодам, амплитудам и температуре: звезды без околозвездной пыли имеют меньшие периоды и амплитуды и они горячее. Их авторы обозначили «голубыми» SRb-звездами. «Красные» SRb имеют темп потери массы, светимости и начальные массы на ГП сравнимые с такими же миридами, но их эффективные температуры незначительно выше. Их периоды содержат первый обертон пульсации.

Углеродные звезды и звезды с технецием найдены только среди «красных» SRb. Практически все циркониевые полуправильные переменные тоже находятся в группе «красных». Кроме того, при аппроксимации инфракрасных спектров чернотельным излучением показано, что только в случае «голубых» кислородных полуправильных переменных возможна интерпретация одним «черным телом», в то время как для остальных групп полуправильных необходимо два [6].

На то, что полуправильные переменные «перемешаны» со звездами типа Миры Кита, указывают многие аспекты. Например, ряд работ по определению зависимости «период – светимость» обнаруживают интересный факт. Авторы Беддинг и Зийлстра [45] на основании зависимости в К-полосе, полученной Фистом [46] для мирид и выведенной ими самими для полуправильных, обнаружили следующее. Те SR-звезды, которые имеют основной период и вторичный, на зависимость «период – светимость» для мирид ложатся с основным периодом, а согласно вторичному периоду эти звезды ложатся на полученную ими зависимость для полуправильных переменных.

6. *Заключение.* Основные решаемые в настоящее время проблемы для полуправильных звезд в целом, те же, что и для звезд типа Миры Кита – построение динамических моделей атмосфер, определение моды пульсации, изучение механизма потери массы, дальнейшая эволюция звезд. Остается еще немаловажный вопрос: трансформация правильных колебаний в хаотические и наоборот. Последняя проблема усложняет изучение этих звезд.

Но, несмотря на сильные нелинейные эффекты (переключения моды пульсации, изменения амплитуды и формы кривой блеска, асимметричные пылевые оболочки) все методы анализа фотометрического поведения, применяемые для звезд типа Миры, можно применят и к полуправильным переменным. Это еще раз подтверждает то, что существуют объекты, которые по фотометрическому поведению относятся одновременно и к миридам и к полуправильным. Например, Y Per (углеродная мирида) на большом интервале времени показывает сложные квазипериодические изменения блеска, типичные для SRb объектов, а сверхгигант S Per (звезда типа SRc) на протяжении нескольких циклов показывала регулярные миридоподобные пульсации с периодом около 816 дней [47], а в последнее время снова стала проявлять хаотические изменения блеска.

Перечислим основные направления работы, которые могут помочь в исследовании процессов пульсаций полуправильных переменных звезд и понимании их эволюции:

- создание атласа средних кривых блеска полуправильных переменных звезд, как например «Catalogue of main characteristics of pulsations of 173 semi-regular stars» [48];
- использование для уточнения классификации звезд трех групп фотометрических параметров (фундаментальных – период $P$, амплитуда $\Delta m = m_{min} - m_{max}$, асимметрия $f = \varphi_{max} - \varphi_{min}$, степень тригонометрического полинома $s$; параметры крутизны ветвей кривой блеска – $m_i = dm(t)/dt$ – максимальный наклон восходящей ветви, $m_d = dm(t)/dt$ – максимальный наклон нисходящей ветви, $t_i = dt/dm$ – характерное время возрастания блеска на $1^m$, $t_d = dt/dm$ – характерное время спадания блеска на $1^m$, $m_{is} = (dm/dt)_{curve}/(dm/dt)_{sinus}$ – для восходящей ветви, $m_{ds}$ – то же самое для нисходящей ветви, где $(dm/dt)_{sinus} = \pi(m_{min} - m_{max})/P$; дополнительные (параметры гармоник) – $r_k$ – амплитуда гармоники с частотой $kf_1$, $\varphi_k$ – фаза максимума гармоники относительно фазы максимума блеска, $\varphi_{k1} = \varphi_k - k\varphi_1$ – сдвиг фаз относительно главного колебания блеска) [36, 49, 50];
- анализ фазовых портретов полуправильных переменных, полученных с различными значениями периодов в случае мультипериодичных звезд [14, 38];
- пересмотр классификации отдельных объектов и выделение переходной группы (полуправильные - мириды) [51, 52, 53] и звезд с «переключением» мод колебаний и мультипериодичностью [54].

Одесский национальный морской университет,
Украина, e-mail: kuda2003@ukr.net